\documentclass[12pt, amsfonts, amssymb]{article}

\usepackage{amssymb}
\usepackage{amsmath}
\usepackage{latexsym}

\newcommand{\CC}{{\mathbb C}}
\newcommand{\RR}{{\mathbb R}}

\newcommand{\PP}{{\mathbb P}}

\renewcommand{\SS}{{\mathbb S}}

\newcommand{\ra}{\rightarrow}

\newcommand{\eps}{\epsilon}
\newcommand{\tQ}{{\tilde Q}}
\newcommand{\tq}{{\tilde q}}
\newcommand{\tpsi}{{\tilde \psi}}
\newcommand{\tA}{{\tilde A}}
\newcommand{\bD}{{\bar D}}

\newcommand{\btheta}{{\bar \theta}}
\newcommand{\bpsi}{{\bar \psi}}
\newcommand{\bxi}{{\bar \xi}}
\newcommand{\blambda}{{\bar \lambda}}
\renewcommand{\L}{{\mathcal L}}
\newcommand{\F}{{\mathcal F}}
\newcommand{\q}{{\mathsf q}}
\newcommand{\Q}{{\mathsf Q}}
\newcommand{\n}{{N_V}}
\renewcommand{\H}{{\mathsf H}}
\renewcommand{\P}{{\mathsf P}}
\newcommand{\M}{{\mathsf M}}
\renewcommand{\S}{{\mathsf S}}

\title{\bf \Large Monopole Operators and Mirror Symmetry in Three Dimensions}
\author{Vadim Borokhov\thanks{borokhov@theory.caltech.edu}, Anton Kapustin\thanks{kapustin@theory.caltech.edu}, Xinkai Wu\thanks{xinkaiwu@theory.caltech.edu}\\
\it California Institute of Technology, Pasadena, CA 91125, USA
}

\begin{document}
\begin{titlepage}

\renewcommand{\thepage}{ }

\maketitle

\begin{abstract}

We study vortex-creating, or monopole, operators in 3d CFTs which
are the infrared limit of $N=2$ and $N=4$ supersymmetric QEDs in
three dimensions. Using large-$N_f$ expansion, we construct monopole 
operators which are primaries of short representations of the superconformal
algebra. Mirror symmetry in three dimensions makes a number of predictions about such operators, and our results confirm these predictions. Furthermore,
we argue that some of our large-$N_f$ results are exact. This implies, in particular, that certain monopole operators in $N=4$ $d=3$ SQED with $N_f=1$ are free fields. This amounts to a proof of 3d mirror symmetry in a special case.

\end{abstract}

\vspace{-6.5in}

\parbox{\linewidth}
{\small\hfill CALT-68-2397}

\end{titlepage}
%\pagestyle{empty}

%\tableofcontents

\section{Introduction}

One of the most remarkable exact results in quantum field theory is the
equivalence of the quantum sine-Gordon model and the massive Thirring
model~\cite{Coleman,Mandelstam}. The ``duality'' between these two
theories has a very transparent physical meaning. Quantum sine-Gordon 
theory contains topological solitons (kinks). It turns out that
a certain operator which has non-zero matrix elements between the vacuum
and the one-kink sector is a fermion and satisfies the equations of
motion of the massive Thirring model~\cite{Mandelstam}. Thus
the duality arises from ``rewriting'' the sine-Gordon model in terms
of kink variables. 

In the last two decades a large number of dualities have been proposed for
quantum field theories in higher dimensions. The first successful
proposal of this sort is the S-duality of $N=4$ $d=4$ super-Yang-Mills
theory~\cite{MO,WO,Osborn}. It is believed that many of these conjectural dualities have the same origin as the sine-Gordon/Thirring duality,
i.e. they arise from ``rewriting'' a theory in terms of new fields
which create topological solitons. But so far nobody managed to
prove a non-trivial higher-dimensional duality along the lines of
Ref.~\cite{Mandelstam}. The main reason for this is that the conjectured
dualities in higher dimensions typically involve non-Abelian
gauge theories and are vastly more complicated than the sine-Gordon/Thirring duality. Usually, it is not even clear which solitons
are ``responsible'' for the duality.

In this paper we report a progress in proving a non-perturbative duality
in three dimensions. This duality, known as 3d mirror symmetry,
has been proposed by K.~Intriligator and N.~Seiberg~\cite{IS}, and
later studied by a number of authors \cite{deBHOO}-\cite{GT}.
Mirror symmetry in three dimensions has a number of special features
that make it more amenable to study than other higher-dimensional
dualities. First of all, mirror symmetry makes sense for Abelian gauge
theories, for which the complications due to the presence of unphysical 
degrees of freedom are not so severe. Second, it is known how to
construct a mirror theory (in fact, many mirror theories~\cite{KS})
for any Abelian gauge theory~\cite{deBHOOY,KS}. The mirror is always an 
Abelian gauge theory, but usually with a different gauge group.
Third, all mirror pairs can be derived from
a certain ``basic'' mirror pair by formal manipulations~\cite{KS}.
This basic example identifies the infrared limit of $N_f=1$ $N=4$ $d=3$ 
SQED with a {\it free} theory of a twisted hypermultiplet. To prove
this basic example of mirror symmetry, one only needs to construct a 
twisted hypermultiplet field out of the fields of $N=4$ SQED and show 
that it is free. Fourth, it is known what the relevant topological soliton 
is in this case: it is none other than the Abrikosov-Nielsen-Olesen 
vortex~\cite{five}.

In our previous paper~\cite{BKWone}, we showed how to define
vortex-creating (or monopole) operators in the infrared limit
of 3d abelian gauge theories. The main tools used were radial 
quantization and large-$N_f$ expansion. 
The only example considered in Ref.~\cite{BKWone} was ordinary
(non-supersymmetric) QED.
In this theory monopole operators have irrational dimensions
at large $N_f$ and do not satisfy any nice equation of motion. 
In this paper we study monopole operators in $N=2$ and $N=4$ SQEDs.
More precisely, we construct monopole operators in 3d SCFTs which
are the infrared limit of $N=2$ and $N=4$ SQEDs. We focus on
operators which live in short multiplets of the superconformal algebra.
The dimensions of primaries of such multiplets saturate a BPS-like
bound, so we will sometimes refer to operators in short multiplets
as BPS operators.

Mirror symmetry makes predictions about the spectrum and other properties 
of BPS operators, including those with non-zero vortex charge. 
In Ref.~\cite{five} some of these predictions have been
verified on the Coulomb branch of $N=2$ SQED, where the infrared theory is free. Our computations are performed at the origin of the moduli space, where the infrared theory is an interacting SCFT. Thus the agreement between our 
results and the predictions of mirror symmetry is a new check of this
duality. In addition, we have been able to verify certain interesting
relations in the chiral ring which follow from mirror symmetry.
In the approach of Ref.~\cite{five}, the origin of these relations was
obscure.

In many cases one can go further and argue that certain results derived at large $N_f$ remain valid even for $N_f$ of order one. For example, our
monopole operators have ``anomalous'' transformation laws under global symmetries, whose form is fixed by quasi-topological considerations 
(the Atiyah-Patodi-Singer index theorem). 
This implies that the global charges of monopole operators do not receive corrections at any order in $1/N_f$ expansion. 
Furthermore, since our monopole operators belong
to short representations of the superconformal algebra, their scaling
dimensions are determined by their transformation law under R-symmetry.
In the case of $N=4$ SQED, where it is easy to identify the relevant R-symmetry, this allows us to determine the exact scaling dimensions of monopole operators for all $N_f$. Our main assumption is that the
$1/N_f$ expansion has a large enough domain of convergence.

If we consider the special case of $N=4$ SQED with $N_f=1$, then the 
above arguments tell us that 
a certain monopole operator is a (twisted) hypermultiplet whose lowest
component is a scalar of dimension $1/2$. In a unitary theory, this is only
possible if the hypermultiplet is free. Thus we are able to show
that for $N_f=1$ certain monopole operators satisfy free equations of motion.
This is essentially the statement of mirror symmetry in this 
particular case. 

The paper is organized as follows. In Section~\ref{sec:ntwo} we study monopole
operators in the infrared limit of $N=2$ $d=3$ SQED at large $N_f$ and 
compare with the predictions of mirror symmetry. In Section~\ref{sec:nfour} 
we do the same for $N=4$ $d=3$ SQED. In Section~\ref{sec:finitenf} we show that certain large-$N_f$ results are exact, and argue that this implies the ``basic'' example of $N=4$ mirror symmetry. 
In Section~\ref{sec:discussion} we discuss our results and list open problems.

\section{Monopole operators in $N=2$ $d=3$ SQED}\label{sec:ntwo}

\subsection{Review of $N=2$ SQED and $N=2$ mirror symmetry}

$N=2$ $d=3$ SQED can be obtained by the dimensional reduction of $N=1$ $d=4$
SQED. The supersymmetry algebra contains a complex spinor supercharge $\Q_\alpha$ and its complex-conjugate $\bar{\Q}_\alpha$.
The field content is the following: a vector multiplet with gauge
group $U(1)$, $N_f$ chiral multiplets of charge $1$ and $N_f$ chiral multiplets
of charge $-1$. We will use $N=2$ superspace to describe these fields.
General superfields are functions of $x\in\RR^{2,1}$, a complex spinor 
$\theta_\alpha,$ and its complex-conjugate $\btheta_\alpha$. 
The vector multiplet is described
by a real superfield $V(x,\theta,\btheta)$ satisfying $V^\dag=V$. 
The corresponding field-strength multiplet is 
$\Sigma=\eps^{\alpha\beta}D_\alpha \bD_\beta V.$ The lowest component of
$\Sigma$ is a real scalar $\chi$, while its top component is the gauge 
field-strength $F_{\mu\nu}$.
The vector multiplet also contains a complex spinor $\lambda_\alpha$ (photino).
A chiral multiplet is described by a superfield $Q(x,\theta,\btheta)$ 
satisfying the chirality constraint:
$$
\bD_\alpha Q=0.
$$
It contains a complex scalar $A,$ a complex spinor $\psi_\alpha,$ and a
complex auxiliary field $F$.
We will denote the superfields describing charge $1$ matter multiplets 
by $Q_j,$ $j=1,\ldots,N_f,$ and the superfields describing charge $-1$ matter multiplets by $\tQ_j,$ $j=1,\ldots,N_f.$ Then the action takes the
form
$$
S_{N=2}=\int d^3 x\ d^4\theta\left\{\frac{1}{e^2}\Sigma^\dag\Sigma+
\sum_{j=1}^{N_f}\left(Q_j^\dag e^V Q_j+\tQ_j^\dag e^{-V} \tQ_j \right)\right\}.
$$
Besides being supersymmetric, this action has a global $SU(N_f)\times
SU(N_f)\times U(1)_B\times U(1)_N$ symmetry. 
The action of $SU(N_f)\times SU(N_f)$ is obvious (it is a remnant of the chiral flavor symmetry of $N=1$ $d=4$ SQED). Under $U(1)_B$ the fields $Q_j$ and
$\tQ_j$ have charges $1$, while $V$ transforms trivially. Finally, there is an R-symmetry $U(1)_N$ under which the fields transform as follows:
\begin{align*}
Q_j(x,\theta,\btheta)&\mapsto 
Q_j\left(x,e^{i\alpha}\theta,e^{-i\alpha}\btheta\right),\\
\tQ_j(x,\theta,\btheta)&\mapsto 
\tQ_j\left(x,e^{i\alpha}\theta,e^{-i\alpha}\btheta\right),\\
V(x,\theta,\btheta)&\mapsto 
V\left(x,e^{i\alpha}\theta,e^{-i\alpha}\btheta\right).
\end{align*}
There is one other conserved current: 
$$
J^\mu=\frac{1}{4\pi} \eps^{\mu\nu\rho} F_{\nu\rho}.
$$ 
Its conservation equivalent to the Bianchi identity.
We will call the corresponding charge the vortex charge, and
the corresponding symmetry $U(1)_J$ symmetry. 
All the fundamental fields have zero vortex charge; our task in this
paper will be to construct operators with non-zero vortex charge
and compute their quantum numbers. Operators with non-zero vortex
charge will be called monopole operators.

One can add an $N=2$ Chern-Simons term to the action of $N=2$ SQED. However,
the theory is consistent without it, and in this paper we will limit
ourselves to the case of vanishing Chern-Simons coupling. 

$N=2$ $d=3$ SQED is super-renormalizable and becomes free in the ultraviolet
limit. In the infrared it flows to an interacting superconformal field
theory (SCFT). Note that the action needs no counter-terms,
if one uses a regularization preserving all the symmetries. Thus the infrared
limit is equivalent to the limit $e\ra\infty$.

In general, the infrared CFT is strongly coupled and quite hard to study.
A simplification arises in the large $N_f$ limit, where the infrared
theory becomes approximately Gaussian. The reason for this is the same as in
the non-supersymmetric case~\cite{BKWone}. At leading order in the large $N_f$
expansion, the matter fields retain their UV dimensions. The dimension
of the gauge field strength multiplet 
$\Sigma$ is $1$ to all orders in $1/N_f$ expansion. This can be traced to the fact that the dual of the gauge field strength is an identically conserved current, as well as a primary field in the infrared 
SCFT.\footnote{In the UV the dual of the field strength is not a primary, but a descendant of a scalar known as the dual photon.} A well-known
theorem states that in a unitary CFT in $d$ dimensions a conserved primary
current has dimension $d-1$. Since the gauge field strength occurs as the
top component of $\Sigma$, and $\theta,\btheta$ have dimension $-1/2$, this
implies that the photino has infrared dimension $3/2$, while the lowest
component $\chi$ has dimension $1$.

The IR dimensions of $Q$ and $\tQ$ can be computed order by order in
$1/N_f$ expansion, but the exact answer for all $N_f$ is unknown. 
The only other thing we know about these dimensions is that they are equal
to the R-charges of $Q$ and $\tQ$. This is a consequence
of the fact that $Q$ and $\tQ$ live in short representation of the superconformal algebra, and therefore their scaling dimensions are 
constrained by unitarity.\footnote{Strictly speaking, it is the dimension 
of gauge-invariant chiral primaries like $Q\tQ$ that is constrained 
by unitarity to be equal to the R-charge. However, since $Q$ and $\tQ$ are chiral superfields, the dimension and R-charge of $Q\tQ$ is twice the dimension and R-charge of $Q$ and $\tQ$, and the claimed result follows.}
However, the R-current in
question is not necessarily the one discussed above. Rather, it is
some unknown linear combination of the $U(1)_N$ and $U(1)_B$ currents. 
We will call it the ``infrared'' R-current, to avoid confusion with 
$U(1)_N$ current defined above. 
In the large $N_f$ limit it is easy to see that the infrared R-charge is
$$
R_{IR}=N+B\left(\frac{1}{2}+O\left(\frac{1}{N_f}\right)\right),
$$
where $N$ and $B$ are the charges corresponding to $U(1)_N$ and $U(1)_B$. 
For $N_f$ of order $1$ we do not know the coefficient in front of $B$, and so cannot easily determine the infrared dimensions of $Q$ and $\tQ$.

For $N_f=1$ mirror symmetry comes to our rescue. The statement of 3d mirror
symmetry in this case is that the IR limit of $N=2$ SQED is the same
as the IR limit of another $N=2$ gauge theory. This other gauge theory
has gauge group $U(1)^{N_f}/U(1)_{diag}$, and $3N_f$ chiral matter multiplets $q_j,\tq_j,S_j,$ $j=1,\ldots,N_f.$ The action of the mirror theory has
the form
\begin{multline*}
S_{dual}=\int d^3x\ d^4\theta \sum_{j=1}^{N_f}\left\{
\frac{1}{e^2}\Sigma_j^\dag\Sigma_j + \frac{1}{e^2} S_j^\dag S_j +
q_j^\dag e^{V_{j}-V_{j-1}} q_j+\tq_j^\dag e^{-V_{j}+V_{j-1}} \tq_j \right\}\\
+\left(\int d^3x\ d^2\theta \sum_{j=1}^{N_f} q_j \tq_j S_j +h.c.\right), 
\end{multline*}
where the gauge multiplets satisfy the constraints
\begin{equation}\label{constr}
V_0=V_{N_f},\quad \sum_{j=1}^{N_f} V_j=0.
\end{equation}
Note that the chiral fields $S_j$ are neutral with respect to the gauge
group and couple to the rest of the theory only through a superpotential.

The mirror theory also flows to a strongly coupled SCFT in the infrared
limit $e\ra\infty$, and in general the mirror description does not help to 
compute the IR scaling dimensions in the original theory. 
However, the case $N_f=1$ is very special: the mirror gauge group becomes
trivial, and the mirror theory reduces to the
Wess-Zumino model in three dimensions with the action
$$
S_{WZ}=\int d^3x\ d^4\theta\left(q^\dag q+ \tq^\dag\tq + S^\dag S\right)
+\left(\int d^3x\ d^2\theta\ q\tq S + h.c.\right).
$$
This theory has ``accidental'' $S_3$ symmetry permuting $q,\tq,$ and
$S,$ which allows one to determine their infrared R-charges. Indeed, since
in the infrared limit the superpotential term must have R-charge $2$,
the R-charges of $q,\tq$ and $S$ must be $2/3$. The mirror map identifies $S$ with the operator $Q\tQ$ in the original theory~\cite{five}.
Thus we infer that for $N_f=1$ $Q$ and $\tQ$ have infrared R-charge $1/3$. 
Comparing with large-$N_f$ results, we see that the infrared R-charge has a non-trivial dependence on $N_f$. 

Let us describe in more detail the matching of global symmetries between
the original and mirror theories following Ref.~\cite{five}. 
The symmetry $U(1)_B$ of the original
theory is mapped to the symmetry under which all $S_j$ have charge $2$, while $q_j$ and $\tq_j$ have charges $-1$. The symmetry $U(1)_J$ is mapped to 
the $U(1)$ symmetry under which all $q_j$ have charge $1/N_f$, all $\tq_j$
have charge $-1/N_f$, while $S_j$ are uncharged. The R-symmetry $U(1)_N$
maps to an R-symmetry under which all $q_j$ and $\tq_j$ have charge $1$ 
and $S_j$ are uncharged. The mapping of non-Abelian
symmetries is not well understood. It is only known that that the currents
corresponding to the Cartan subalgebra of the diagonal $SU(N_f)$ are mapped 
to the $N_f-1$ $U(1)_J$ currents of the mirror theory. 

\subsection{Monopole operators in $N=2$ SQED at large $N_f$}

Our strategy for studying monopole operators will be the same as in
Ref.~\cite{BKWone}.
In any 3d conformal field theory, there is a one-to-one map between
local operators on $\RR^3$ and normalizable states of the same theory on
$\SS^2\times\RR$. Therefore we will look for states with non-zero vortex
charge on $\SS^2\times\RR$. In other words, we will be studying
$N=2$ SQED on $\SS^2\times\RR$ in the presence of a
magnetic flux on $\SS^2.$ 
Since our goal is to check the predictions of
mirror symmetry, we will require that the states be annihilated by half of
the supercharges; then the corresponding local operators will live in
short representations of the superconformal algebra. The low-energy
limit of $N=2$ SQED is an interacting SCFT, so in order to make computations
possible, we will take $N_f$ to be very large. This has the effect of making
the CFT weakly coupled. In particular, in the large $N_f$ limit the fluctuations
of the gauge field and its superpartners are suppressed, and one can treat
them as a classical background. In other words, at leading order in $1/N_f$
we end up with free chiral superfields coupled to an appropriate background
vector superfield. We will discuss how one can go beyond the large-$N_f$
approximation in Section~\ref{sec:finitenf}.

The states on $\SS^2\times\RR$ of interest to us
are in some sense BPS-saturated, since they are annihilated by half of 
the supercharges. But in contrast to the situation in flat space, here the
supercharges do not commute with the Hamiltonian $\H$ which generates
translations on $\RR$. Indeed, since the Hamiltonian on $\SS^2\times\RR$ is the same as the dilatation generator on $\RR^3$, and supercharges have 
dimension $1/2$, it follows that the supercharges obey
$$
[\Q_\alpha,\H]=-\frac{1}{2} \Q_\alpha, \quad [{\bar\Q}_\alpha,\H]=-\frac{1}{2}{\bar\Q}_\alpha.
$$ 
Note also that in the radial quantization approach $\Q_\alpha$ and
${\bar\Q}_\alpha$
are no longer Hermitian conjugate of each other. Rather, their Hermitian
conjugates are superconformal boosts $\S_\alpha$ and ${\bar\S}_\alpha$, which
have dimension $-1/2$.

For the same reasons as in Ref.~\cite{BKWone},
in the large $N_f$ limit the energy $E$ of the states with non-zero 
vortex charge is of order $N_f$. By unitarity, for scalar states
$E$ is bounded from below by the R-charge $R_{IR}$.
Furthermore, we will see below that in the limit $N_f\ra\infty$ $R_{IR}$
is also of order $N_f$, while the combination $E-R_{IR}$ stays finite for 
all the states we encounter. A similar limit in $d=4$ SCFTs recently gained
some prominence in connection with AdS/CFT correspondence~\cite{BMN}. But
unlike Ref.~\cite{BMN}, we take the number of flavors, rather than the 
number of colors, to infinity.

First let us determine which classical background on $\SS^2\times\RR$ we need
to consider. As in Ref.~\cite{BKWone}, we have a gauge field on 
$\SS^2\times\RR$
with a magnetic flux $n$. Assuming rotational invariance of the 
large-$N_f$ saddle point, this implies that we have a constant magnetic 
field on $\SS^2$. The only other bosonic field in the $N=2$ vector multiplet 
is the real scalar $\chi$. It is determined by the condition of the vanishing of the photino variation under half of the SUSY transformations. This will
ensure that the monopole operator we are constructing is a chiral primary.

It is convenient to work out the photino variations on $\RR^3,$ and then 
make a conformal transformation to $\SS^2\times\RR$. 
Photino variations in Euclidean
$N=2$ SQED on $\RR^3$ have the form
\begin{align*}
\delta\lambda &=i\left(-\sigma^i\partial_i \chi+\frac{1}{2}\eps^{ijk}\sigma^k 
F_{ij}+D\right)\xi,\\
\delta\blambda &=i\left(-\sigma^i\partial_i \chi-\frac{1}{2}\eps^{ijk}\sigma^k
F_{ij}- D\right)\bxi,
\end{align*}
where $\xi$ and $\bxi$ are complex spinors which parametrize SUSY variations.
(In Euclidean signature, they are not related by complex conjugation.)
Since we are setting the background values of the matter fields to zero,
the D-term can be dropped.
Half-BPS states are annihilated by $\bxi_\alpha{\bar\Q}^\alpha$ for any $\bxi$ 
and therefore must satisfy
$$
F=-*d\chi.
$$
Hence the scalar background on $\RR^3$ is
$$
\chi=\frac{n}{2r},
$$
where $n$ is the vortex charge (the magnetic charge of the Dirac monopole 
on $\RR^3$).
Unsurprisingly, supersymmetry requires the bosonic field configuration to be 
an abelian BPS monopole.
Recalling that $\chi$ has dimension $1$ in the infrared, we infer that
on $\SS^2$ the scalar background is simply a constant:
$$
\chi=\frac{n}{2}.
$$
Similarly, an anti-BPS state is annihilated by $\xi_\alpha\Q^\alpha$ for 
any $\xi$, and therefore the scalar field on $\SS^2$ is
$$
\chi=-\frac{n}{2}.
$$

Having fixed the classical background, we proceed to compute the spectrum of matter field fluctuations. The details of the computation are explained in
the Appendix. The results are as follows. The energy spectrum of charged 
scalars is the same for $A^j$ and $\tA^j$, does not
depend on whether one is dealing with a BPS or an anti-BPS configuration, and
is given by
$$
E=\pm E_p=\pm\left(\frac{|n|-1}{2}+p\right),\quad p=1,2,\ldots.
$$
The degeneracy of the $p^{\rm th}$ eigenvalue is $2|E_p|$, and the
corresponding eigenfunctions transform as an irreducible representation of the
rotation group $SU(2)_{rot}$. The spectrum is symmetric with respect to 
$E\ra -E$. 

The energy spectrum of charged spinors is the same for $\psi^j$ and $\tpsi^j$
and is given by
\begin{align*}
E&=E_p^+=\frac{|n|}{2}+p,\quad p=1,2,\ldots,\\
E&=E_p^-=-\frac{|n|}{2}-p,\quad p=1,2,\ldots,\\
E&=E_0=\pm \frac{|n|}{2}.
\end{align*}
Here the upper (lower) sign refers to the BPS (anti-BPS) configuration. 
The eigenspace with eigenvalue $E$ has degeneracy $2|E|$ and furnishes
an irreducible representation of $SU(2)_{rot}$.

Comparing the fermion spectrum with the results of Ref.~\cite{BKWone}, we see
that the inclusion of the scalar $\chi$ causes dramatic changes in the spectrum of fermions. First, unlike in Ref.~\cite{BKWone}, there are no zero modes. 
Second, the spectrum is not symmetric with respect to $E\ra -E$.

The absence of zero modes, either in the scalar or in the spinor sector,
means that for a fixed magnetic flux the state of lowest energy is unique. 
We will call it the vacuum state. By construction, it is an (anti-) BPS state, 
and we would like to determine its quantum numbers. It is clear that the vacuum
state is rotationally invariant, so its spin is zero. It is also
a flavor singlet. The other quantum numbers
of interest are the energy (which is the same as the conformal dimension of
the corresponding local operator~\cite{BKWone}) and the $U(1)_B$ and 
$U(1)_N$ charges.
Vacuum energy and charge are plagued by normal-ordering ambiguities,
as usual, but as in Ref.~\cite{BKWone} we can deal with them by requiring the
state corresponding to the unit operator (i.e. the vacuum with zero magnetic
flux) to have zero energy and charges. 

The asymmetry of the fermionic energy spectrum leads to a subtlety in the
computation.
Suppose we use point-splitting regularization to define vacuum energy and
charges. Then one gets different results after renormalization depending
on the ordering of operators $\psi$ and $\bpsi$. For example, consider 
two definitions of the $U(1)_N$ charge
\begin{align*}
N(\tau)&=\lim_{\beta\ra 0+} \left[\int_{\SS^2} 
-\bpsi\left(\tau+\frac{\beta}{2}\right)\sigma_\tau 
\psi\left(\tau-\frac{\beta}{2}\right) - 
{\bar\tpsi}\left(\tau+\frac{\beta}{2}\right)\sigma_\tau 
\tpsi\left(\tau-\frac{\beta}{2}\right)\right.\\
&\left. -C(\beta)\right],\\ 
N'(\tau)&=\lim_{\beta\ra 0+} \left[\int_{\SS^2} \psi\left(\tau+\frac{\beta}{2}\right)\sigma_\tau
\bpsi(\tau-\frac{\beta}{2})
+\tpsi\left(\tau+\frac{\beta}{2}\right)\sigma_\tau
{\bar\tpsi}\left(\tau-\frac{\beta}{2}\right)\right.\\ 
&\left. -C'(\beta)\right],
\end{align*}
where $\tau$ is the time coordinate on $\SS^2\times\RR$, and $C(\beta)$ and
$C'(\beta)$ are c-numbers defined as the $U(1)_N$ charge of the vacuum with $n=0$ regularized by means of appropriate point-splitting. 
One can easily 
see that these two definitions are equivalent only if the fermion spectrum
is symmetric with respect to zero; otherwise they differ by a c-number
which depends on $n$. This ambiguity can be removed by requiring that
the regularization procedure preserve charge-conjugation symmetry. 
This mandates using expressions symmetrized with respect to $\psi$ and $\bpsi$
(and $\tpsi$ and $\bar{\tpsi}$).
Thus we will define the $U(1)_N$ charge as the average of $N(\tau)$ and
$N'(\tau)$. The same applies to the $U(1)_B$ charge and the energy operator. 

As an illustration, let us compute the $U(1)_N$ charge of the vacuum for
arbitrary $n$.
The above definition yields the following regularized $U(1)_N$ charge:
\begin{equation}\label{Nreg}
N_{reg}(\beta)=N_f \sum_{E} 2|E|\, {\rm sign}(E)\, e^{-\beta \vert E\vert}.
\end{equation}
Here the summation extends over the fermion energy spectrum, and
we took into account that $\psi$ and $\tpsi$ have the same spectrum and
$U(1)_N$ charge and contribute equally to $N_{reg}(\beta).$
The regularized charge of the unit operator is identically zero,
since the spectrum is symmetric for $n=0$. For non-zero vortex charge the spectrum is symmetric except for a single eigenvalue $E_0$. 
Thus the renormalized charge is equal to 
$$
N_{vac}=\pm \lim_{\beta\ra 0+} N_f |n|=\pm N_f|n|,
$$
where the upper (lower) sign refers to the BPS (anti-BPS) state.
Since the spectrum of scalars is symmetric, only spinors will contribute
to the $U(1)_B$ charge of the vacuum, and an identical argument gives
$$
B_{vac}=\mp N_f|n|.
$$
A similar, but slightly longer, computation gives the vacuum energy:
$$
E=\frac{|n|N_f}{2}.
$$
This is the same as the scaling dimension of the corresponding monopole
operator.

Recall that at large $N_f$ the R-charge which is the superpartner of the
Hamiltonian is given by
$$
R_{IR}=N+\frac{1}{2}B.
$$
It is easy to see from the above results that $E=\pm R_{IR}$ 
for our ``vacuum'' states. This is a satisfying result, since in a unitary 
3d CFT the scaling dimension of any (anti-) chiral primary must be equal to 
(minus) its R-charge.

As expected, the energy and the R-charge of the vacuum are of order
$N_f$. Other states can be obtained by acting on the vacuum with a finite
number of creation operators for the charged fields. If the number of
creation operators is kept fixed in the limit of large $N_f$, then
both $E$ and $R_{IR}$ tend to infinity, with $E-R_{IR}$ kept finite. Thus the limit we are considering is qualitatively similar to the PP-wave limit of
$N=4$ $d=4$ SYM theory considered in Ref.~\cite{BMN}. But since we are
taking the number of flavors, rather than the number of colors, to
infinity, the physics is rather different. For example, in Ref.~\cite{BMN}
the combination $R^2/N_c$ is kept fixed and can be an arbitrary positive real number (it is the effective string coupling in the dual string theory). 
The analogous quantity in our case is $2R_{IR}/N_f=|n|$, the vortex charge, 
which is quantized.

One issue which we have not mentioned yet is gauge-invariance.
In order for the operator to be gauge-invariant, the corresponding state
must satisfy the Gauss law constraint. In the limit $e\ra\infty$
this is equivalent to requiring that the state be annihilated by
the electric charge density operator~\cite{BKWone}. For the vacuum state, 
this is automatic.
For excited states, the Gauss law constraint is a non-trivial requirement.

We have identified above a scalar state on $\SS^2\times \RR$ which is
a chiral primary. What about its superpartners? The key point is to realize
that the classical field configuration we are considering breaks some
of the symmetries of the CFT. In such a situation, one must enlarge the
Hilbert space by extra variables (``zero modes'') which correspond to
the broken generators. In other words, the semi-classical Hilbert space is
obtained by tensoring the ``naive'' Hilbert space by the space of functions
on the coset $G/H$, where $G$ is the symmetry group of the theory, and
$H$ is the invariance subgroup of the classical configuration. This observation
plays an important role in the quantization of solitons. For example, if we
are dealing with a soliton in a Poincar\'{e}-invariant theory which breaks translational symmetry to nothing, but preserves rotational symmetry, the zero
mode Hilbert space is 
$$
ISO(d-1,1)/SO(d-1,1)=\RR^{d-1,1}.
$$
Poincare group acts on the space of functions on $\RR^{d-1,1}$ in the usual
manner. Furthermore, if a soliton breaks some of supersymmetries, there will
be fermionic zero modes, and the bosonic coset must be replaced by an
appropriate supercoset. 

In our case, the symmetry of theory is described by the $N=2$ $d=3$ super-Poincare group.\footnote{We may forget about $U(1)_N$, $U(1)_B$, and
the flavor symmetry, since they are left unbroken by our field configuration.
Furthermore, although conformal and superconformal boosts do not preserve 
our field configuration, they can be ignored, since these symmetry generators cannot be exponentiated to well-defined symmetry transformations on $\RR^3.$}
For the BPS state, the invariance subgroup is generated by rotations
and the complex supercharge ${\bar\Q}_\alpha$.
Thus the zero mode Hilbert space will consist of functions on the supercoset
$$
\frac{\left\{\M_{ij}, \P_i, \Q_\alpha, 
{\bar\Q}_\alpha\right\}}{\left\{\M_{ij}, {\bar\Q}_\alpha\right\}},
$$
where $\M_{ij}$ and $\P_i$ are the rotation and translation generators on $\RR^3$, respectively, and $\{ A,B,\ldots\}$ denotes the super-group with Lie
super-algebra spanned by $A,B,\ldots.$
Functions on this supercoset are nothing but $N=2$ $d=3$ chiral
superfields~\cite{HS}. Thus the usual rules of semi-classical 
quantization lead to the conclusion that the BPS monopole operator is 
described by a chiral superfield. Similarly, an anti-BPS monopole operator 
will be described by an anti-chiral superfield. In particular, $N=2$ 
auxiliary fields are automatically incorporated. 
(Note that at large $N_f$ our monopole operators are not expected to satisfy 
any closed equation of motion. On the other hand, auxiliary fields can
be eliminated only on-shell. This suggests that any description of
monopole operators without auxiliary fields would be rather cumbersome.)

\subsection{A comparison with the predictions of $N=2$ mirror symmetry}

As explained above, under mirror symmetry the vortex charge is mapped to 
$1/N_f$ times the charge which ``counts'' the number of $q$'s minus the 
number of $\tq$'s. Thus the obvious gauge-invariant chiral primaries with
vortex charge $\pm 1$ are
$$
V_+=q_1 q_2\ldots q_{N_f},\quad V_-=\tq_1 \tq_2\ldots \tq_{N_f}.
$$
Using the matching of global symmetries explained above, we see that both
$V_+$ and $V_-$ are singlets under $SU(N_f)\times SU(N_f)$ flavor symmetry, have $U(1)_B$ charge $-N_f$ and $U(1)_N$ charge $N_f$. 
Comparing this with the previous subsection, we see that $V_+$ has
the same quantum numbers as the BPS state
with $n=1$ that we have found, while $V_-^\dagger$ has the same quantum 
numbers as the anti-BPS state with $n=1$.
This agreement provides a non-trivial check of $N=2$ mirror symmetry.

Our computation of the charges was performed in the large-$N_f$ limit,
but mirror symmetry predicts that the result remains true for $N_f$ of
order $1$. Can we understand this apparent lack of $1/N_f$ corrections
to $U(1)_N$ and $U(1)_B$ charges?
The answer is yes: $U(1)_N$ and $U(1)_B$ charges are not corrected at any
order in $1/N_f$ expansion
because they can be determined by quasi-topological considerations 
($L^2$ index theorem on $\SS^2\times\RR$).
This will be discussed in more detail in Section~\ref{sec:finitenf}.

\section{Monopole operators in $N=4$ $d=3$ SQED}\label{sec:nfour}

\subsection{Review of $N=4$ SQED and $N=4$ mirror symmetry}

$N=4$ $d=3$ SQED is the dimensional reduction of $N=2$ $d=4$ SQED. The
supersymmetry algebra includes two complex spinor supercharges
$\Q^i_\alpha,$ $i=1,2$ and their complex conjugates. In Minkowski signature,
the spinor representation is real, so we may also say that we have four
real spinor supercharges. 
If we regard $N=4$ SQED as an $N=2$ $d=3$ gauge theory, then it contains, besides the fields of $N=2$ SQED, a chiral superfield $\Phi$. 
This superfield is neutral and together with the $N=2$ vector multiplet $V$ forms an $N=4$ vector multiplet. The chiral superfields $Q_j$ and $\tQ_j^\dag$
combine into an $N=4$ hypermultiplet. The action of $N=4$ SQED is the sum 
of the action of $N=2$ SQED, the usual kinetic term for $\Phi$,
and a superpotential term
$$
\int d^3x\, d^2\theta \sum_{j=1}^{N_f} Q_j\Phi \tQ_j + h.c.
$$ 
The flavor symmetry of this theory is $SU(N_f)$. In addition, there
is an important R-symmetry $SU(2)_R\times SU(2)_N$. In the $N=2$ superfield
formalism used above, only its maximal torus $U(1)^2$ is manifest. 
The lowest components of $Q$ and $\tQ^\dag$ are singlets under $SU(2)_N$
and transform as a doublet under $SU(2)_R$. The complex scalar $\Phi$ in the
chiral multiplet and the real scalar $\chi$ in the $N=2$ vector multiplet transform as a triplet of $SU(2)_N$ and are singlets of $SU(2)_R$. 
The transformation properties of other fields can be inferred from these using the fact that the four real spinor supercharges of $N=4$ SQED transform in the $(2,2)$ representation of $SU(2)_R\times SU(2)_N$. 

Although there is a complete symmetry between $SU(2)_R$ and $SU(2)_N$ at
the level of superalgebra, the transformation properties of
fields do not respect this symmetry. Therefore one can define twisted
vector multiplets and twisted hypermultiplets for which the roles of
$SU(2)_N$ and $SU(2)_R$ are reversed. $N=4$ SQED contains only
``ordinary'' vector and hypermultiplets, while its mirror (see below)
contains only twisted multiplets. There are interesting $N=4$ theories in 3d
which include both kinds of multiplets~\cite{BG,KS}, but in this paper we
will only consider the traditional ones, which can be obtained by dimensional
reduction from $N=2$ $d=4$ theories.

In order to make contact with our discussion of
$N=2$ SQED, we will denote the global $U(1)$ symmetry under which
$Q$ and $\tQ$ have charge $1$ and $\Phi$ has charge $-2$ by $U(1)_B$,
and we will denote an R-symmetry under which $Q$ and $\tQ$ are neutral
and $\Phi$ has charge $2$ by $U(1)_N$. It is easy to see that $U(1)_N$
is a maximal torus of $SU(2)_N$, while the generator of $U(1)_B$
is a linear combination of the generators of $SU(2)_N$ and $SU(2)_R$.
The generator of the maximal torus of $SU(2)_R$ can be taken as
$$
R=N+B.
$$

$N=4$ SQED is free in the UV and flows to an interacting SCFT in the IR.
The infrared dimensions of fields in short multiplets of the superconformal
algebra are determined by their spin and transformation properties under
$SU(2)_R\times SU(2)_N$.  This is easily seen in the harmonic superspace
formalism, where the compatibility of constraints on the superfields leads to
relations between the dimension and the R-spins~\cite{HS}. For gauge-invariant
operators, one can alternatively use arguments based on unitarity (see
e.g. Ref.~\cite{Minwalla}). 

Perhaps the easiest way to work out the relation
between the IR dimension and $SU(2)_R\times SU(2)_N$ quantum numbers
is to regard $N=4$ SQED as a special kind of $N=2$ theory. That is, it
is an $N=2$ gauge theory which has, besides a manifest complex supercharge,
a non-manifest one. It is easy to
see that the combination $N+\frac{1}{2} B$ is the generator of
the $U(1)$ subgroup of $SU(2)_N\times SU(2)_R$ with respect to which 
the manifest supercharge has charge $1$, 
while the non-manifest supercharge has charge $0$. In the IR limit, the corresponding current is in the
same multiplet as the stress-energy tensor (because all $SU(2)_R\times SU(2)_N$
currents are), and therefore the dimension of chiral primary states
must be equal to their charges with respect to $N+\frac{1}{2}B$. (Note
that in the case of $N=2$ SQED this was true only in the large-$N_f$
limit.) In particular, the IR dimensions of $Q_j$ and $\tQ_j$ are
$1/2$, and the IR dimension of $\Phi$ and $\chi$ is $1$. 

According to Ref.~\cite{IS}, the mirror theory for $N=4$ SQED is a (twisted)
$N=4$ $d=3$ gauge theory with gauge group $U(1)^{N_f}/U(1)_{diag}$ and $N_f$ (twisted) hypermultiplets $(q_j,\tq_j)$. The matter multiplets
transform under the gauge group as follows:
$$
q_j\ra q_j e^{i(\alpha_j-\alpha_{j-1})}, \quad 
\tq_j\ra \tq_j e^{-i(\alpha_j-\alpha_{j-1})},\quad j=1,\ldots,N_f,
$$
where we set $\alpha_0=\alpha_{N_f}$. 
The action of the mirror theory is
\begin{multline*}
S_{dual}=\int d^3x\ d^4\theta \sum_{j=1}^{N_f}\left\{
\frac{1}{e^2}\Sigma_j^\dag\Sigma_j + \frac{1}{e^2} S_j^\dag S_j+
q_j^\dag e^{V_{j}-V_{j-1}} q_j+
\tq_j^\dag e^{-V_{j}+V_{j-1}} \tq_j\right\}\\
+\left(\int d^3x\ d^2\theta \sum_{j=1}^{N_f} q_j \tq_j (S_j-S_{j-1})
+h.c.\right), 
\end{multline*}
Here $N=2$ vector multiplets $V_j$ satisfy the constraints Eq.~(\ref{constr}), 
$N=2$ chiral multiplets $S_j$ satisfy similar constraints
$$
S_0=S_{N_f},\quad \sum_{j=1}^{N_f} S_j=0,
$$
and each pair $(V_j,S_j)$ forms a (twisted) $N=4$ vector multiplet.

The matching of global symmetries goes as follows. The R-symmetries are
trivially identified. The vortex current of $N=4$ SQED is mapped to $1/N_f$
times the Noether current corresponding to the following global $U(1)$
symmetry:
$$
q_j\ra e^{i\alpha} q_j,\quad \tq_j\ra e^{-i\alpha}\tq_j,\quad j=1,\ldots,N_f.
$$
The currents corresponding to the maximal torus of $SU(N_f)$ flavor symmetry
of $N=4$ SQED are mapped to the vortex currents
$$
2\pi\, J_j=*F_j,\ j=1,\ldots,N_f,\quad \sum_{j=1}^{N_f} J_j=0,
$$
where $F_j$ is the field-strength of the $j^{\rm th}$ gauge field.
The mapping of the rest of $SU(N_f)$ currents is not well understood.

\subsection{Monopole operators in $N=4$ SQED at large $N_f$}

To begin with, we can regard $N=4$ SQED as a rather special $N=2$ gauge
theory, and look for BPS and anti-BPS monopole operators in this theory.
This amounts to focusing on a particular $N=2$ subalgebra of the
$N=4$ superalgebra. Different choices of an $N=2$ subalgebra are all related
by an $SU(2)_N$ transformation, so we do not loose anything by doing this.

>From this point of view, our problem is almost exactly the same as in the
case of $N=2$ SQED. The only difference between the two is the presence
of the chiral superfield $\Phi$. But in the large $N_f$ limit it becomes
non-dynamical, and the $N=2$  BPS condition requires the background value of
$\Phi$ to be zero. This implies that the radial quantization of the matter
fields $Q_j,\tQ_j$ proceeds in {\it exactly} the same way as in the
$N=2$ case and yields the same answer for the spectrum and properties
of BPS and anti-BPS states. Namely, for any vortex charge $n$ we have a 
single BPS and a single anti-BPS states, with charges
$$
N=\pm |n| N_f,\quad B=\mp |n|N_f,
$$
and energy $E=|n|N_f/2.$

An interesting new element in the $N=4$ case is the way short
multiplets of $N=2$ superconformal symmetry fit into a short multiplet
of $N=4$ superconformal symmetry. Recall that we have made a certain
choice of $N=2$ subalgebra of the $N=4$ superalgebra. 
This choice is preserved by the $U(1)_N$ symmetry, but not by the 
$SU(2)_N$ symmetry. Thus we have an $SU(2)/U(1)\simeq \CC\PP^1$ worth of BPS conditions. Applying an $SU(2)_N$ rotation to the BPS state found above, we obtain a half-BPS state for every point on $\CC\PP^1$. These half-BPS states
fit into a line bundle $\L$ over $\CC\PP^1$. Similarly, applying $SU(2)_N$
transformations to the anti-BPS state, we obtain another line bundle
on $\CC\PP^1$ which is obviously the complex conjugate of $\L$.

The $\CC\PP^1$ which parametrizes different choices of the $N=2$
subalgebra has a very clear meaning in the large $N_f$ limit. Namely,
we chose the scalar background on $\SS^2\times\RR$ to be 
$\Phi=0,\chi=\frac{n}{2}$, but obviously any $SU(2)_N$
transform of this is also a half-BPS configuration. The manifold
of possible scalar backgrounds is a 2-sphere given by
$$
|\Phi|^2+\chi^2=\left(\frac{n}{2}\right)^2.
$$
The BPS state we
are interested in is the Fock vacuum of charged
matter fields on $\SS^2\times\RR$ in a {\it fixed} background. 
As we vary the background
values of $\Phi$ and $\chi$, we obtain a bundle of Fock vacua on 
$\SS^2\sim\CC\PP^1.$ This bundle can be non-trivial because of Berry's phase~\cite{Berry,Simon}.

Now we can easily see how $N=4$ superconformal symmetry is realized
in our formalism. As argued above, we need to enlarge our Hilbert
space by the Hilbert space of zero modes, which arise because the
classical background breaks some of the symmetries of the theory.
Compared to the $N=2$ case, we have additional bosonic zero modes
coming from the breaking of R-symmetry from $SU(2)_N$ down to $U(1)_N$.
Thus our fields will depend on coordinates on $\RR^3\times \CC\PP^1$.
As for fermionic zero modes, in the BPS case they are generated by a complex
spinor supercharge which depends on the coordinates on $\CC\PP^1$
as follows:
$$
\Q_\alpha=\sum_{i=1,2} u_i \Q^i_\alpha.
$$
Here $u_1,u_2\in \CC$ are homogeneous coordinates on $\CC\PP^1$,
and $\Q^i_\alpha,$ $i=1,2$ are a pair of complex spinor supercharges
which transform as a doublet of $SU(2)_N$. Therefore monopole
operators will be described by ``functions'' on the
supermanifold
$$
S(\RR^3)\boxtimes {\mathcal O}(1),
$$
where $S(\RR^3)$ is the trivial spinor bundle on 
$\RR^3$ (with fiber coordinates regarded as Grassmann-odd), while 
${\mathcal O}(1)$ is the tautological line bundle 
on $\CC\PP^1$. We put the word ``functions'' in quotes, because, as explained above, we may need to consider sections of non-trivial line bundles on $\CC\PP^1$ instead of functions.

This supermanifold is known as the {\it analytic superspace}~\cite{HSorig1,HSorig2,HS} (see also
Section 3 of Ref.~\cite{Z}).
It is a chiral version of the so-called harmonic superspace. 
It is well known that ``functions'' on the analytic superspace (analytic
superfields) furnish short representations of the superconformal algebra 
with eight supercharges~\cite{HS}. We conclude that in the large-$N_f$ 
limit BPS monopole operators are described by $N=4$ $d=3$ analytic superfields.
Needless to say, anti-BPS monopole operators are described by anti-analytic
superfields which are complex-conjugates of the analytic ones.

It remains to pin down the topology of the bundle $\L$
over $\CC\PP^1$. Since this is a line bundle, its topology is completely
characterized by the first Chern class. A ``cheap'' way to find the Chern
class is to note
that the scaling dimension of an analytic superfield
(more precisely, of its scalar component) is equal to half the Chern
number of the corresponding line bundle. (The Chern number is the value
of the first Chern class on the fundamental homology class of $\CC\PP^1.$) 
This follows from the way superconformal algebra is represented on analytic 
superfields~\cite{HS}. We already know the dimension of our BPS state,
and therefore infer that the Chern number of $\L$ is equal to $N_f|n|$.

We can also determine the Chern number directly, by computing the curvature
of the Berry connection for the bundle of Fock vacua. In the present case,
the computation is almost trivial, since the Hamiltonians at different points
of $\CC\PP^1$ are related by an $SU(2)_N$ transformation. In particular,
it is sufficient to compute the curvature at any point on $\CC\PP^1$.
For example, we can identify $\CC\PP^1$ with a unit sphere in $\RR^3$
with coordinates $(x,y,z)$ and compute the curvature at the
``North Pole,'' which has Euclidean coordinates $(0,0,1)$. (The abstract
coordinates $(x,y,z)$ can be identified with $({\rm Re\,\Phi},
{\rm Im}\,\Phi,\chi)$.)
Using $SU(2)_N$ invariance,
we easily see that the Fock vacuum at the point $(x,y,z)$ with $z\simeq 1$,
$x,y\ll 1$ is given by
$$
\vert x,y,z\rangle=\exp\left(i\left(\frac{x}{z} N_x-\frac{y}{z} N_y\right)+
O(x^2+y^2)\right)\vert 0,0,1\rangle.
$$
Here $N_x$ and $N_y$ are the generators of $SU(2)_N$ rotations about
$x$ and $y$ axes. Therefore the curvature of the Berry connection 
at the point $(0,0,1)$ is
\begin{multline*}
\F=i\left(d\vert x,y,z\rangle, \wedge d\vert x,y,z\rangle\right)=
i dx\wedge dy \langle 0,0,1\vert [N_y,N_x]\vert 0,0,1\rangle\\
=dx\wedge dy \langle 0,0,1\vert N_z\vert 0,0,1\rangle.
\end{multline*}
Now we recall that the vacuum at $(0,0,1)$ is an eigenstate of $N_z$
with eigenvalue $\pm N_f|n|/2$ (one needs to remember that $N=2N_z$).
Taking into account that $\F$ is 
an $SU(2)_N$-invariant 2-form on $\CC\PP^1$, we conclude that it is
given by
$$
\F=\pm\frac{1}{2}N_f|n|\, \Omega,
$$
where $\Omega$ is the volume form on the unit 2-sphere. It follows that
the Chern number of the Fock vacuum bundle is
$$
c_1=\frac{1}{2\pi} \int_{S^2} \F=\pm N_f|n|,
$$
where the upper (lower) sign refers to $\L$ (resp. $\L^*$).
The result agrees with the indirect argument given above.

\subsection{A comparison with the predictions of $N=4$ mirror symmetry}

Chiral primaries in the mirror theory with vortex number $\pm 1$ are 
exactly the same as in the $N=2$ case, i.e.
$$
V_+=q_1 q_2\ldots q_{N_f}, \quad V_-=\tq_1\tq_2\ldots \tq_{N_f}.
$$
Their $U(1)_N$ and $U(1)_B$ quantum numbers match those computed in
the original theory using radial quantization and large-$N_f$ expansion.
This provides a check of $N=4$ mirror symmetry at the origin of the
moduli space. We can also translate this into the language of analytic
superfields. Then a hypermultiplet $(q_j,\tq_j^\dag)$ is described
by an analytic superfield $\q_j$ whose Chern number is $1$ (in the notation
of Ref.~\cite{HS}, it would be written as $q^+_j$, where a single $+$
superscript refers to the unit Chern number). The analytic superfield which
is gauge-invariant and carries vortex charge $1$ is given by
$$
\q_1\q_2\ldots \q_{N_f}.
$$
It has Chern number $N_f$, and in the notation of Ref.~\cite{HS} it
would have $N_f$ superscripts. This field corresponds to the
BPS multiplet constructed in the previous section, while its complex
conjugate corresponds to the anti-BPS multiplet.

Mirror symmetry also predicts a certain interesting relation in the
chiral ring of the IR limit of $N=4$ SQED. 
Consider the product of $V_+$ and $V_-$: 
$$
V_+V_-=(q_1\tq_1)(q_2\tq_2)\ldots (q_{N_f}\tq_{N_f}). 
$$ 
Using the equation of motion for $S_j$, it is easy to see that the 
operators $(q_j \tq_j)$ for different $j$ are equal modulo descendants. Furthermore, mirror symmetry maps any of these operators to $\Phi$ modulo descendants~\cite{five}. Thus we infer that modulo descendants we have a relation in the chiral ring:
\begin{equation}\label{rel}
V_+V_-\sim \Phi^{N_f}.
\end{equation}
Can we understand this relation in terms of $N=4$ SQED? Indeed we can!

To begin with, it is easy to see that the operator $\Phi^{N_f}$ is the
only chiral operator whose quantum numbers match those of $V_+V_-$
and which could appear in the OPE of $V_+$ and $V_-$.
Thus it is sufficient to demonstrate that it appears with a non-zero
coefficient. To this end, we need to compute the 3-point function
of $V_+,V_-,$ and $\left(\Phi^\dag\right)^{N_f}.$ In the radial
quantization approach, we need to show that the matrix element
$$
\langle V_-^\dag \vert \left(\Phi^\dag\right)^{N_f} \vert V_+\rangle
$$
is non-zero.

Now we recall that the state corresponding to $V_+$ has magnetic flux
$+1$ and scalar VEV $\chi=\frac{1}{2}$, while the state corresponding to
$V_-$ has magnetic flux $-1$ and $\chi=-\frac{1}{2}.$
Hermitian conjugation reverses the sign of the magnetic flux and leaves the 
VEV of $\chi$ unchanged. It follows that the path integral which computes
the matrix element of any operator between $\langle V_-^\dag\vert $ and 
$\vert V_+\rangle$ must be performed over field configurations such
that the magnetic flux is equal to $1$, while the scalar $\chi$
asymptotes to $1/2$ at $\tau=-\infty$ and $-1/2$ at $\tau=+\infty.$
Thus we are dealing with a kink on $\SS^2\times\RR$.

Next, we note that the Dirac operator on $\SS^2\times\RR$ coupled
to such a background may very well have normalizable zero modes. If this 
is the case, then in order to get a non-zero matrix element one needs
to insert an operator which has the right quantum numbers to absorb
the zero modes. For example, one can insert a product of all fermionic
fields which possess a zero mode. Another possibility, which is more
relevant for us, is to insert some bosonic fields which interact with fermions
and can absorb the zero modes. In our case, the action contains a complex 
scalar $\Phi$ which has Yukawa interactions of the form
$$
\int d^3x\, \Phi\sum_{j=1}^{N_f} \psi_j\tpsi_j.
$$
Thus if each $\psi$ and each $\tpsi$ has a single normalizable zero
mode, then we can get a non-zero result for the matrix element if we insert
precisely $N_f$ powers of $\Phi^\dag.$

To complete the argument it remains to show that the Dirac operator for both
$\psi$ and $\tpsi$ has a single zero mode. The Atiyah-Patodi-Singer
theorem says in this case that the $L^2$ index of the Dirac operator is
$$
ind(D)=\frac{1}{2}(\eta(H_-)-\eta(H_+)),
$$
where $\eta(H_\pm)$ denotes the $\eta$-invariant of the asymptotic
Dirac Hamiltonian at $\tau\ra \pm \infty.$ We also made use of the fact
that neither $H_+$ nor $H_-$ have zero modes (see Section~\ref{sec:ntwo}). 
Now we recall that we have computed the $\eta$-invariants already:
according to Eq.~(\ref{Nreg}), $\eta(H_-)$ and $\eta(H_+)$ coincide with the $U(1)_N$ charges of the BPS and anti-BPS vacua, respectively, 
divided by $N_f$. This implies that the index of the Dirac operator is 
equal to $1$, for both $\psi$ and $\tpsi,$ and therefore both $\psi$ 
and $\tpsi$ have a single zero mode.

\section{Beyond the large-$N_f$ limit}\label{sec:finitenf}

\subsection{Non-renormalization theorems for the anomalous charges}

We have seen that mirror symmetry makes certain predictions about
the quantum numbers of BPS monopole operators, and that our large-$N_f$
computations confirm these predictions. But mirror symmetry also
suggests that large-$N_f$ results for $U(1)_B$ and $U(1)_N$ charges
remain valid for all $N_f$, all the way down to $N_f=1$.
In this subsection we provide an explanation for this without appealing
to mirror symmetry. We show that the values of $U(1)_N$ and $U(1)_B$
charges for monopole operators are fixed by the $L^2$ index
theorem for the Dirac operator on $\SS^2\times\RR$ and therefore cannot
receive $1/N_f$ corrections. 

The argument is very simple. For concreteness, consider the monopole
operators $V_\pm$ which have vortex charge $n=\pm 1$. These operators
are related by charge conjugation and thus have the same $U(1)_N$ charge,
which we denote $\n$. To determine $\n$, we need to consider the
transition amplitude on $\SS^2\times\RR$ from the state corresponding to 
$V_+$ to the state corresponding to $V_-^\dag$: if it violates the 
$U(1)_N$ charge by $m$, then $\n=-m/2$. Since $\psi$ and $\tpsi$ have
$N=-1$, the charge is violated by $-2N_f$ times the index of the Dirac
operator on $\SS^2\times\RR$. The index of the Dirac operator in the
present case has only boundary contributions ($\eta$-invariants),
which depend on the asymptotics of the gauge field and the scalar $\chi.$
When these asymptotics are given by the large-$N_f$ saddle points,
the index was evaluated in Section~\ref{sec:nfour} with the result
$ind(D)=1.$ Furthermore, in the large-$N_f$ expansion fluctuations
about the saddle point are treated using perturbation theory.
Hence to all orders in $1/N_f$ expansion the transition amplitude
from $V_+$ to $V_-^\dag$ will violate $U(1)_N$ charge by $-2N_f$.
This implies that the $U(1)_N$ charge of $V_\pm$ is equal to $N_f$ to
all orders in $1/N_f$ expansion. An identical argument can be made for
$U(1)_B$.

One may ask if it is possible to dispense with the crutch of $1/N_f$ expansion
altogether. Naively, there is no problem: we consider the path
integral for $N=4$ or $N=2$ SQED with $e=\infty$ and use the
APS index theorem to infer the charges of $V_\pm.$ However, this
argument is only formal, because we do not know how to make sense
of this path integral without using $1/N_f$ expansion. In particular,
this leads to difficulties with the evaluation of the index: we cannot compute
the $\eta$-invariants without knowing the precise asymptotic form of the
background, but the asymptotic conditions put constraints only on
the total magnetic flux through $\SS^2$ and the average value of
$\chi$ at $\tau=\pm\infty.$ (We remind that the $L^2$-index of a Dirac
operator on a non-compact manifold is only a quasi-topological quantity,
which can change if the asymptotic behavior of the fields is changed.)
The index has a definite value only if we choose some particular
asymptotics for the gauge field and $\chi$.

\subsection{A derivation of the basic $N=4$ mirror symmetry}

It is plausible that the point $N_f=1$ is within the radius of convergence
of $1/N_f$ expansion. Singularities in an expansion parameter usually
signal some sort of phase transition, and in the case of $N=4$ SQED we do not
expect any drastic change of behavior as one decreases $N_f$.

With this assumption, we can prove the basic example of $N=4$ mirror
symmetry, namely, that the IR limit of $N=4$ SQED with $N_f=1$ is
dual to the theory of a free twisted hypermultiplet. The proof is quite
straightforward. As explained above, the $U(1)_N$ charge
of the chiral field $V_\pm$ is equal to $N_f$ to all orders in 
$1/N_f$ expansion, while its $U(1)_B$ charge is equal to $-N_f$.
This implies that the IR dimension of $V_+$ is equal to $N_f/2$ to 
all orders in $1/N_f$ expansion (see Section~\ref{sec:nfour}). 
Assuming that $1/N_f$ expansion converges at $N_f=1$, this implies
that for $N_f=1$ the IR dimension of $V_\pm$ is $1/2$. In a unitary 3d CFT,
a scalar of dimension $1/2$ must be free~\cite{Minwalla}. Then, by virtue
of supersymmetry, the $N=2$ superfields $V_\pm$
are free chiral superfields with $N=1$ and $B=-1$, or, equivalently,
the pair $(V_+,V_-^\dag)$ is a free twisted hypermultiplet.

The above argument shows that the IR limit of $N=4$ SQED contains a free
sector generated by the action of free fields $V_\pm$ on the vacuum.
But this sector also contains all the states generated by $\Phi$ and its superpartners. Indeed, the product of $V_+$ and $V_-$ is a chiral field
which has zero vortex charge and $N=2,B=-2.$ It is easy to see that
the only such field is $\Phi$. In addition, since $V_+$ and $V_-$ are
independent free fields, their product is non-zero. Thus we must
have $V_+V_-\sim \Phi$ (we have seen above how a more general relation 
Eq.~(\ref{rel}) can be demonstrated in the large-$N_f$ limit). 
We conclude that the sector of the IR limit of $N=4$ SQED generated by 
$\Phi$ and its superpartners is contained in the charge-0 sector of the 
theory of a free twisted hypermultiplet. This is precisely the statement of
mirror symmetry in this particular case.
 
\section{Discussion of results and open problems}\label{sec:discussion}

In this paper we showed that many predictions of three-dimensional
mirror symmetry can be verified directly at the origin of the moduli
space, where the IR theory is an interacting SCFT. The main idea was that
the $e\ra\infty$ limit of 3d gauge theories can be defined in the
continuum using large-$N_f$ expansion, and then vortex-creating operators
can be rigorously defined as well. Focusing on vortex-creating operators
in short representations of the superconformal algebra, we showed that
their transformation laws under various symmetries are determined by
index theorems on $\SS^2\times\RR$ and therefore are not corrected at
any order in $1/N_f.$ In the $N=4$ case, this allowed us to determine
the exact scaling dimensions of vortex-creating operators to all orders
in $1/N_f$ expansion. If we assume that $N_f=1$ is within the convergence
radius of this expansion, we can prove the basic $N=4$ mirror symmetry,
which says that a certain large sector of the IR limit of $N=4$ $N_f=1$
SQED can be described in terms of a free twisted hypermultiplet.

We feel that these results go some way towards making the 3d mirror symmetry
conjecture into a theorem (on the physical level of rigor). 
On the other hand, much yet remains to be
done before one can claim that one understands 3d mirror symmetry.
First, it would be desirable to construct monopole operators
directly, using Hamiltonian formalism on $\RR^3$, rather than by
identifying the corresponding states on $\SS^2\times\RR.$ Mandelstam's
construction of soliton-creating operators in the sine-Gordon 
theory~\cite{Mandelstam} serves as a model in this respect. Second,
it would be interesting to find the mirror of more
complicated observables in $N=4$ SQED. Third, mirror symmetry predicts
that many 3d gauge theories have ``accidental'' symmetries in the infrared
limit~\cite{IS,Kapustin}. It appears possible to understand the origin of 
these symmetries using the methods of this paper.
Fourth, for $N_f>1$ the mirror theory of $N=4$ SQED is a gauge theory,
and one would like to have a conceptual understanding
of the origin of the dual gauge group. Although all Abelian mirror pairs
can be derived for the ``basic'' one, the derivation
is rather formal and does not shed much light on this question. 

More ambitiously, we would like to extend the approach of this paper to
non-Abelian gauge theories and non-Abelian 3d mirror symmetry. It seems
that a pre-requisite for this is the ability to construct 
operators which are not invariant with respect to the dual gauge group
out of the original variables (i.e. construct operators representing
``dual electrons'' or ``dual quarks.'') This problem is also
the major stumbling block for our understanding of 4d dualities, and we
hope that studying 3d mirror symmetry will eventually lead to a progress
in proving 4d dualities.

\section{Appendix: radial quantization of $N=2$ SQED}

We start with the Lagrangian of $N=1$ $d=4$ SQED in the conventions
of Wess and Bagger~\cite{WB} and perform a Wick rotation to Euclidean
signature:
$$
{\cal L}_{\RR^4}=\left.-{\cal L}_{\RR^{1,3}}\right|_{x^0=-it}, \quad \left.V_0\right|_{\RR^{1,3}}= \left.i\chi\right|_{\RR^4},
$$
where $V_0$ is the time-like coordinate of the $U(1)$ connection.
Then we require that all fields be independent of the Euclidean
time $t$. This procedure gives the action density for
$N=2$ $d=3$ SQED on Euclidean $\RR^3$:
\begin{align*}
{\cal L}=&i\bar{\psi}\vec{\sigma}(\vec{\nabla}+i\vec{V})\psi+i\chi\bar{\psi}
\psi+
i\bar{\tilde{\psi}}\vec{\sigma}(\vec{\nabla}-i\vec{V})\tilde{\psi}-i\chi\bar{
\tilde{\psi}}\tilde{\psi}+\chi^2\left(AA^*+\tilde{A}\tilde{A}^*\right)\\
&+([\vec{\nabla}+i\vec{V}]A)([\vec{\nabla}-i\vec{V}]A^*)+
([\vec{\nabla}-i\vec{V}]\tilde{A})([\vec{\nabla}+i\vec{V}]\tilde{A}^*)\\
&-D(AA^*-\tilde{A}\tilde{A}^*)
+i\sqrt{2}(A\bar{\psi}\bar{\lambda}-A^*\psi\lambda-
\tilde{A}\bar{\tilde{\psi}}\bar{\lambda}+\tilde{A}^*\tilde{\psi}\lambda)+
O\left(\frac{1}{e^2}\right).
\end{align*}
In the infrared limit $e\to\infty$ the kinetic terms for the vector multiplet
can be ignored. Note also that in the $e\ra\infty$ limit the equation of
motion for $D$ enforces the vanishing of D-terms.

To go from $\RR^3$ to $\SS^2\times \RR$, we perform
a Weyl rescaling of the Euclidean metric $ds^2=dr^2+r^2 d\Omega^2$ by a 
factor $1/r^2$. If we set $r=e^\tau$, then $\tau$ is
an affine parameter on $\RR$. The component fields of $Q$ must be rescaled 
as follows:
$$
\psi\ra e^{-\tau}\psi,\quad \bar{\psi}\ra e^{-\tau}\bar{\psi}, \quad
A\ra e^{-\frac{\tau}{2}}A,\quad A^*\ra e^{-\frac{\tau}{2}}A^*.
$$
The component fields of $\tQ$ transform in a similar way. 
The bosonic fields in the vector multiplet transform as follows:
$$
\chi \to e^{-\tau}\chi,\quad \vec{V}\to\vec{V}.
$$

To find the one-particle energy spectrum for charged fields, we use the
procedure and notations of Ref.~\cite{BKWone}.
The Lagrangian for $\psi$ and $\bar{\psi}$ in the background of the
(anti-) BPS monopole on $\RR^3$ has the following form 
$$
{\cal L}[\psi,\bar{\psi}]_{\SS^2\times\RR}=i\bar{\psi}\sigma_r
\left[{\frac{\partial}{\partial\tau}}-\left(\vec{J}^2-\vec{L}^2+
\frac14\right)-q\sigma_r \mp q\sigma_r\right]\psi,
$$
where $q=-eg=-n/2,$ and the upper (lower) sign corresponds to a BPS 
(anti-BPS) monopole. A solution with energy
$E$ has the form $\psi\sim e^{-E\tau}$, $\bar{\psi}\sim e^{E\tau}$.
The above Lagrangian is the same as in Ref.~\cite{BKWone}, except
for the last term in brackets. We will not repeat the diagonalization
procedure and simply quote the resulting energy spectrum for 
$\psi$ and $\tilde{\psi}$:
$$
-\frac{|n|}{2}-p,\quad \mp\frac{|n|}{2},\quad \frac{|n|}{2}+p,
$$
where $p=1,2,\ldots.$
%Whereas the energy spectrum for $\bar{\psi}$ and $\bar{\tilde{\psi}}$ is
%$$
%E_p=-\frac{|n|}{2}-p,\quad \pm\frac{|n|}{2},\qquad, \frac{|n|}{2}+p. 
%$$
Each energy-level has spin $j=|E|-1/2$ and degeneracy $2j+1=2|E|$.

The Lagrangian for $A$, $A^*$ is
$$
{\cal L}[A,A^*]_{\SS^2\times\RR}=[(\vec{\nabla}_a+i\vec{V}_a)A]
[(\vec{\nabla}_b-i\vec{V}_b)A^*]g^{ab}+\frac14 AA^*+\chi^2 AA^*.
$$
The equation of motion for $A$ has the from 
$$
\frac{d^2}{d\tau^2}A=\left(\vec{L}^2+\frac14\right)A,
$$
where $\vec{L}$ is the generalized angular momentum defined in Ref.~\cite{WY}.
Using the known spectrum of $\vec{L}^2$, we easily find the one-particle
energy spectrum for $A$ and $\tilde{A}$:
$$
-\frac{|n|-1}{2}-p,\quad \frac{|n|-1}{2}+p,\quad p=1,2,\ldots.
$$
The degeneracy of each eigenvalue is again $2|E|$, and each eigenspace
is an irreducible representation of the rotation group.

\section*{Acknowledgments}
We would like to thank Jim Gates,  Takuya Okuda, Hiroshi Ooguri,
John Preskill, and Mark Wise for discussions. A.K. is also grateful 
to Matt Strassler for numerous conversations during the years of 1998 and 1999 which contributed to the genesis of this paper. This work was supported in part by a DOE grant DE-FG03-92-ER40701.

\end{document}